\documentclass[5p]{elsarticle}

\usepackage{amssymb,amsmath,amstext}
\usepackage{mathtools, cuted}
\usepackage[utf8]{inputenc}
\usepackage[english]{babel}
\usepackage[T1]{fontenc}

\usepackage{bm}
\usepackage{color}
\usepackage{hyperref}

\usepackage{amsxtra}

\newcommand{\R}{\mathbb R}
\newcommand{\CC}{\mathbb{C}}

\newcommand{\vc}[1]{\boldsymbol{#1}}
\newcommand{\vt}[1]{\mathsf{#1}}
\newcommand{\tsp}{\mathsf{T}}

\newcommand{\tr}{\operatorname{\mathrm{tr}}}

\newcommand{\E}{{\vt{E}}}
\newcommand{\ee}{{\vc{\varepsilon}}}
\newcommand{\T}{{\vt{T}}}
\newcommand{\ttau}{{\vc{\tau}}}

\newcommand{\F}{{\vt{F}}}

\newcommand{\J}{{\vt{J}}}

\newcommand{\Z}{{\vt{Z}}}
\newcommand{\I}{{\vt{I}}}

\newcommand{\dyad}[2]{{#1}\otimes{#2}}
\newcommand{\gl}{\mathcal{G}_\mathcal{L}}
\newcommand{\gm}{\mathcal{G}_\mathcal{M}}

\title{Periodic rhomboidal cells for symmetry-preserving homogenization and isotropic metamaterials}

\author[1]{Giulio G. Giusteri\corref{cor1}}
\ead{giulio.giusteri@math.unipd.it}

\author[2]{Raimondo Penta}
\ead{Raimondo.Penta@glasgow.ac.uk}

\cortext[cor1]{Corresponding author}

\address[1]{Dipartimento di Matematica``Tullio Levi-Civita'', Universit\`a degli Studi di Padova, via Trieste 63, 35121, Padova, Italy}

\address[2]{School of Mathematics and Statistics, University of Glasgow, University Place, G128QQ, Glasgow, UK}

\begin{document}

\begin{abstract}
In the design and analysis of composite materials based on periodic arrangements of sub-units it is of paramount importance to control the emergent material symmetry in relation to the elastic response.
The target material symmetry plays also an important role in additive manufacturing.
In numerous applications it would be useful to obtain effectively isotropic materials. While these typically emerge from a random microstructure, it is not obvious how to achieve isotropy with a periodic order.
We prove that arrangements of inclusions based on a rhomboidal cell that generates the Face-Centered Cubic lattice do in fact preserve any material symmetry of the constituents, so that spherical inclusions of isotropic materials in an isotropic matrix produce effectively isotropic composites.
\end{abstract}


\maketitle

\section{Introduction}

Composite or microstructured materials have been long since considered as important means to engineer and optimize mechanical properties for specific applications \cite{cherkaev1997topics, milton2002theory, hull1996introduction}. 
With the advent of additive manufacturing (also known as 3D-printing), 
production of artificial constructs conceived to possess specific optimal properties is now becoming possible.
The design of the mechanical behaviour of composites is increasingly relevant in a large variety of scenarios of practical interests, ranging from construction \cite{nicholas2020integrating} to biomimetic materials \cite{suresh2020review}.

The architecture of such materials is typically based on designing features at a small scale, that lead to the desired large-scale behavior of structural elements.
In light of this, theoretical studies of composites often involve asymptotic (periodic) homogenization
or alternative upscaling techniques based on average field theories (see for example the review \cite{HoriNematNasser1999} where the two approaches are compared) to obtain suitable predictions of the effective material behavior. Examples can be found in Refs. \cite{royer2019quasi}, \cite{penta2016can, collis2017multi}, and \cite{nika2019design} concerning poroelastic composites, biophysical applications (such as bone, tendons, tumors, and organs) and metamaterials, respectively.

On the one hand, obtaining detailed quantitative information on material parameters typically requires employment of sophisticated combinations of analytical and/or computational techniques \cite{Berger_2017,Rossi_2020,Yera_2020,Wu_2021,Rossi_2021}.
On the other hand, some qualitative information can be deduced by simple symmetry arguments and this will be the focus of the present note.
One of the most important qualitative properties of elastic solids is the material symmetry, and the target symmetry for a composite is always considered in the design process.
Luckily, one can draw some conclusion on material symmetry by considering how the response to simple homogeneous deformations interacts with the geometric properties of the assembly of microscopic units that form a composite.
This is especially true when the large-scale specimen is built by a periodic reproduction of identical units.

Several studies dealt with material symmetries in the context of linearized elasticity \cite{Forte_1996,Bona_2004} and the minimal symmetry induced by a periodic arrangement of inclusions in a binary composite has been repeatedly investigated by different methods \cite{Ptashnyk_2016,Podesta_2019,Mendez_2019}.
Nevertheless, it is not clear whether it be possible to obtain an effectively isotropic response by a periodic arrangement of inclusions in a three-dimensional body, while it is well known that a hexagonal lattice would suffice to get such a maximal symmetry in a two-dimensional context. 
In particular, a periodic arrangement of uniaxially aligned fibers leads to either tetragonal or transverse isotropic response depending on weather the planar projections of such fibers are encoded in a square or hexagonal periodic cell \cite{sabina2002overall, parnell2006dynamic}.
Topological optimization procedures offer a way to approach an isotropic response by resorting to nontrivial geometries. While in rare cases the lack of isotropy is negligible~\cite{Rossi_2020,Yera_2020}, a residual anisotropy is usually found in three-dimensional composites in the presence periodicity. Introducing randomness in the microstructure remains the most accepted way to reach an isotropic response.

Based on geometric symmetry considerations, it is well known how to achieve a large-scale cubic symmetry with a periodic inclusion lattice.
By considering a standard cubic periodic cell, then the existence of three planes of symmetry guarantees the material orthotropy as long as no additional degree of anisotropy is induced by the material symmetry of the individual phases in the composite. This is true when the individual phases are geometrically arranged in the host in order to guarantee the existence of such planes of symmetries, and they are individually at most orthotropic. When assuming, in addition, that the individual phases are either all isotropic or at most cubic, and that the resulting geometry is invariant with respect to rotation of the three orthogonal axis (this can be achieved for example by considering either a cubic or a spherical inclusion in three dimensions), the resulting material response is then in general cubic. This is also shown in the works \cite{penta2015investigation} and \cite{penta2017asymptotic}, where the authors define a suitable function which measures the deviation from isotropy for composites in the context of asymptotic homogenisation.

Common lore suggests that, with a periodic inclusion lattice, one cannot achieve isotropy without additional constraining strategies \cite{Rossi_2020,Yera_2020}, even with isotropic components.
Should this be the case, then it would be somewhat unpleasant because, on the one hand, periodic arrangements are by far the most convenient approach in both additive manufacturing and computational material design and, on the other hand, isotropic elasticity is very often assumed in practical applications, especially in the context of simple models used to validate experimental results.

In this work, we show that the maximal symmetry that can be achieved for three-dimensional periodic composites is not the cubic one.
We give a rigorous and yet simple proof of the fact that a periodic arrangement on a Face-Centered Cubic (FCC) lattice of spherical inclusions of an isotropic solid within an isotropic matrix gives rise to a large-scale isotropic response.
In doing so, we also show that any rhomboidal computational cell that generates such a lattice can be used to successfully design homogenized solids in which the material symmetry is not affected by the periodicity of the construction, since the latter would preserve even the largest possible symmetry group.
It is significant to observe that the geometric symmetry group of such rhomboidal cells is strictly smaller than the symmetry group of the lattice they generate, but the lattice and not the cell is the geometrically relevant structure when analyzing large-scale properties. 

Incidentally, our result shows that the small lack of isotropy computationally found for several periodic arrangements based on a FCC lattice is to be ascribed to (unavoidable) numerical approximations rather than to real geometric obstructions. 
This leads to important changes in perspective for the interpretation of numerical results and towards the design of isotropically elastic metamaterials, with important consequences on several applications.

We frame our discussion in the context of linear elasticity by introducing, in Section~\ref{sec:Voigt}, a normalized Voigt representation of the elasticity tensor which is very convenient for the identification of material parameters and symmetries. We then discuss the link between lattice symmetries and material symmetries for periodic composites in Section~\ref{sec:geomsym} and, finally, present our main result and a symmetry-preserving rhomboidal cell in Section~\ref{sec:sympcell}.

\section{Normalized Voigt representation}\label{sec:Voigt}

We are interested in describing the effective linear elastic response of a composite that consists of two isotropically elastic phases. 
One is the matrix and the other one occupies spherical inclusions with centers distributed on a periodic lattice. 
Due to the spherical shape of the inclusions and the isotropic nature of the two materials, the only source of anisotropy in the homogenized material response can be the geometry of the inclusion lattice. 
It is thus convenient to represent the linearized measure of strain and the Cauchy stress tensor on a basis for the space of symmetric tensors that is adapted to the geometry of the inclusion lattice. This leads to the construction of a normalized Voigt representation.

We denote by $(\vc a_1,\vc a_2,\vc a_3)$ the \emph{generators} of the lattice, namely linearly independent vectors such that the centers of the spherical inclusions are obtained as combinations of  $\vc a_1$, $\vc a_2$, and $\vc a_3$ with integer coefficients.
The set of lattice sites is then denoted by $\mathcal L=\langle\vc a_1,\vc a_2,\vc a_3\rangle_\mathbb{Z}$.
A set of \emph{directors} of the lattice can be constructed building an orthonormal basis for $\R^3$ out of the generators. For instance, we may choose
\begin{gather*}
\vc l_1=\frac{\vc a_1}{\Vert\vc a_1\Vert},\qquad \Vert \vc l_2\Vert \vc l_2=\vc a_2-(\vc a_2\cdot\vc l_1)\vc l_1,\\
\Vert \vc l_3\Vert \vc l_3=\vc a_3-(\vc a_3\cdot\vc l_1)\vc l_1-(\vc a_3\cdot\vc l_2)\vc l_2.
\end{gather*}

We now introduce an orthogonal basis for the linear space of symmetric tensors built upon the lattice directors. The basis $\mathcal Z=(\Z_1,\Z_2,\Z_3,\Z_4,\Z_5,\Z_6)$ is given in terms of dyadic products by
\begin{gather*}
\Z_1=\dyad{\vc l_1}{\vc l_1}\,,\quad\Z_2=\dyad{\vc l_2}{\vc l_2}\,,\quad\Z_3=\dyad{\vc l_3}{\vc l_3}\,,\\
\Z_4=\frac{\dyad{\vc l_2}{\vc l_3}+\dyad{\vc l_3}{\vc l_2}}{\sqrt{2}}\,,\quad\Z_5=\frac{\dyad{\vc l_1}{\vc l_3}+\dyad{\vc l_3}{\vc l_1}}{\sqrt{2}}
\,,\\
\Z_6=\frac{\dyad{\vc l_1}{\vc l_2}+\dyad{\vc l_2}{\vc l_1}}{\sqrt{2}}\,.
\end{gather*}
Such a basis is orthonormal with respect to the tensor scalar product defined by $\vt A:\vt B:=\tr(\vt A^\tsp\vt B)$.

As customary in linear elasticity, we decompose the deformation gradient tensor as $\F=\J+\nabla\vc u$, with $\vc u$ the displacement field and $\J$ the isometry that maps spatial vectors to material ones.
The standard linearized strain measure is then $\E=\tfrac{1}{2}(\nabla\vc u\J+\J^\tsp\nabla\vc u^\tsp)$. 
In the infinitesimal-displacement regime considered in linear elasticity, $\J$ is constant and homogeneous and can be taken as the identity.
$\E$ is a symmetric tensor, characterized by six degrees of freedom.
A possible choice of objective quantities to represent them are the eigenvalues of $\E$ and the orientation of its eigenvectors with respect to the lattice directors.
Equivalent degrees of freedom can be encoded in the six components of $\E$ on the basis $\mathcal Z$, that we group in a six-component vector $\ee$, so that 
\[
\E=\sum_{k=1}^6\ee_k\Z_k.
\]
The Cauchy stress tensor $\T$ is constitutively related to $\E$ and the lattice directors (all objective quantities). 
To emphasize its relation with the lattice structure, it is convenient to expand also $\T$ on the tensorial basis $\mathcal Z$.
We thus have
\[
\T=\hat{\T}(\E,\vc l_1,\vc l_2,\vc l_3)=\sum_{k=1}^6\ttau_k(\E,\vc l_1,\vc l_2,\vc l_3)\Z_k,
\]
where the six-component vector $\ttau$ corresponding to the normalized Voigt representation of $\T$ has been introduced.
The dependence of the stress components on the lattice directors is written explicitly to highlight the fact that, to describe an objective anisotropic response, we need to consider the strain $\E$ in relation to the lattice directors and not to an arbitrary basis of $\R^3$.

A general linear constitutive relation whereby $\ttau$ depends on the six objective degrees of freedom in $\ee$ is thus given by $\ttau=C\ee$, where $C$ is the 6 by 6 matrix of material coefficients. We can thus write the stress-strain relation as
\begin{equation}\label{eq:constitutive}
\T=\sum_{i,k=1}^6C_{ik}\ee_k\Z_i.
\end{equation}
Since all of the involved quantities are objective, this is a manifestly objective constitutive relation (in line with the conclusion of Steigmann~\cite{Steigmann_2007}).

It is customary to think of the elastic response as generated by a potential energy density.
A necessary and sufficient condition for $\vt T$ in equation \eqref{eq:constitutive} to be the first variation of an elastic energy density quadratic in $\E$ is that the matrix $C$ be symmetric, namely $C_{ik}=C_{ki}$.
This would reduce its independent components from 36 to 21, but this assumption is not necessary for our discussion.
Moreover, the fact that any deformation from the relaxed configuration should increase the stored elastic energy translates into requiring that the matrix of the coefficients $C_{ik}$ be positive definite.

The matrix $C$ is the normalized Voigt representation of the classical elasticity tensor $\CC$ (a fourth order tensor with well-known symmetries) and, indeed, we can also give the constitutive prescription as $\T=\CC\E$.
The definition of $\CC$ implied by relation \eqref{eq:constitutive} is
\[
\CC:=\sum_{i,k=1}^6 C_{ik}\dyad{\Z_i}{\Z_k},
\]
from which we can retrieve the 81 components (greek indices running from 1 to 3)
\[
\CC_{\alpha\beta\gamma\delta}:=\sum_{i,k=1}^6 C_{ik}\dyad{\Z_i^{\alpha\beta}}{\Z_k^{\gamma\delta}}.
\]
We want to emphasize that, while the components of $\CC$ obviously depend on the choice of a basis to compute coordinates in the lab frame, the entries of the elasticity matrix $C$ are material constants completely independent of the reference frame. In fact, they bear a mechanical meaning that is linked only to the mechanical role of the lattice directors chosen to build the tensorial basis $\mathcal Z$.
In view of this, it should be clear that the common pragmatic way of introducing the standard Voigt representation through an index-based identification of the entries of $C$ and $\CC$ confers to the coordinate basis a mechanical meaning that may not always be appropriate.

\section{Lattice symmetries and material symmetries}\label{sec:geomsym}

The symmetries of the lattice are the subgroup of the isometries of $\R^3$ that map the set  $\mathcal L$ of lattice sites onto itself.
Besides the obvious translations that are responsible for the large-scale homogeneity of the material, we can have reflections and rotations and will focus our attention on these.

An important role in our analysis is played by the tensorial basis $\mathcal Z$. It allows to extract only the necessary information from the elasticity tensor in a coordinate-invariant way, thereby giving a material meaning to the coefficients of the 6 by 6 matrix representation $C$.
It provides a consistent way to identify how the components of the stress change when the applied strain is changed by an isometry.
Thanks to the use of the basis $\mathcal Z$, all of the constraints on the coefficients of $C$ derived in what follows are independent of the choice of coordinates used to represent the lattice generators and directors.

Employing index notation with summation only over repeated greek indices going from $1$ to $3$, we have
\begin{gather*}
\Z_k^{\alpha\beta}={\vc l_k}^\alpha{\vc l_k}^\beta\quad\text{for }k=1,2,3,\\
\Z_k^{\alpha\beta}=\frac{{\vc l_{k-1}}^\alpha{\vc l_{k-2}}^\beta+{\vc l_{k-2}}^\alpha{\vc l_{k-1}}^\beta}{\sqrt{2}}\quad\text{for }k=4,5,6, 
\end{gather*}
with pedices computed mod 3.

If we denote by $\vt S$ an isometry, it is represented on the basis $(\vc l_1,\vc l_2,\vc l_3)$ by an orthogonal matrix.
When we apply the isometry $\vt S$ to the basis vectors we get the transformed tensors $\tilde{\vt Z}_k=\vt S\vt Z_k\vt S^\tsp$. More specifically
\[
{\tilde{\vt Z}_k}^{\alpha\beta}=\vt S^{\alpha\mu}{\vc l_k}^\mu\vt S^{\beta\nu}{\vc l_k}^\nu,\quad \text{for }k=1,2,3,\text{ and}
\]
\[
{\tilde{\vt Z}_k}^{\alpha\beta}=\frac{1}{\sqrt{2}}\big(\vt S^{\alpha\mu}{\vc l_{k-1}}^\mu\vt S^{\beta\nu}{\vc l_{k-2}}^\nu+\vt S^{\alpha\mu}{\vc l_{k-2}}^\mu\vt S^{\beta\nu}{\vc l_{k-1}}^\nu\big),
\]
for $k=4,5,6$.
Tensorial scalar products between the basis tensors correspond to summation over the Greek indices. Hence, the products between original (index $j$) and transformed (index $k$) tensors become combinations of the components of the isometry $\vt S$, identified in terms of the Latin indices $j$ and $k$. We thus find 
\begin{itemize}
\item for $j,k=1,2,3$:
\begin{equation}
\Z_j^{\alpha\beta}{\tilde{\vt Z}_k}^{\alpha\beta}={\vc l_j}^\alpha\vt S^{\alpha\mu}{\vc l_k}^\mu\,{\vc l_j}^\beta\vt S^{\beta\nu}{\vc l_k}^\nu=(\vt S^{jk})^2,
\end{equation}
\item for $j=1,2,3$; $k=4,5,6$:
\begin{equation}
\Z_j^{\alpha\beta}{\tilde{\vt Z}_k}^{\alpha\beta}=\sqrt{2} \vt S^{j(k-2)}\vt S^{j(k-1)},
\end{equation}
\item for $j=4,5,6$; $k=1,2,3$:
\begin{equation}
\Z_j^{\alpha\beta}{\tilde{\vt Z}_k}^{\alpha\beta}=\sqrt{2} \vt S^{(j-2)k}\vt S^{(j-1)k},
\end{equation}
\item for $j,k=4,5,6$:
\begin{multline}
\Z_j^{\alpha\beta}{\tilde{\vt Z}_k}^{\alpha\beta}= \vt S^{(j-2)(k-2)}\vt S^{(j-1)(k-1)}\\+\vt S^{(j-2)(k-1)}\vt S^{(j-1)(k-2)}.
\end{multline}
\end{itemize}

From these relations we can write the linear transformation $\hat{S}$ induced by $\vt S$ on the normalized Voigt representation of the strain, namely the matrix representing the change of basis from $(\tilde{\Z}_k)_{k=1}^6$ to $\mathcal Z$:

\begin{strip}
\begin{equation}\label{eq:hatS}
\hat{S}=\begin{pmatrix}
 (\vt S^{11})^2 & (\vt S^{12})^2 & (\vt S^{13})^2 & \sqrt{2} \vt S^{12}\vt S^{13} & \sqrt{2} \vt S^{13}\vt S^{11} & \sqrt{2} \vt S^{11}\vt S^{12} \\
 (\vt S^{21})^2 & (\vt S^{22})^2 & (\vt S^{23})^2 & \sqrt{2} \vt S^{22}\vt S^{23} & \sqrt{2} \vt S^{23}\vt S^{21} & \sqrt{2} \vt S^{21}\vt S^{22} \\
 (\vt S^{31})^2 & (\vt S^{32})^2 & (\vt S^{33})^2 & \sqrt{2} \vt S^{32}\vt S^{33} & \sqrt{2} \vt S^{33}\vt S^{31} & \sqrt{2} \vt S^{31}\vt S^{32} \\
 \sqrt{2} \vt S^{21}\vt S^{31} & \sqrt{2} \vt S^{22}\vt S^{32} & \sqrt{2} \vt S^{23}\vt S^{33} & \vt S^{22}\vt S^{33}+\vt S^{23}\vt S^{32} & \vt S^{23}\vt S^{31}+\vt S^{21}\vt S^{33} & \vt S^{21}\vt S^{32}+\vt S^{22}\vt S^{31} \\
 \sqrt{2} \vt S^{31}\vt S^{11} & \sqrt{2} \vt S^{32}\vt S^{12} & \sqrt{2} \vt S^{33}\vt S^{13} & \vt S^{32}\vt S^{13}+\vt S^{33}\vt S^{12} & \vt S^{33}\vt S^{11}+\vt S^{31}\vt S^{13} & \vt S^{31}\vt S^{12}+\vt S^{32}\vt S^{11} \\
 \sqrt{2} \vt S^{11}\vt S^{21} & \sqrt{2} \vt S^{12}\vt S^{22} & \sqrt{2} \vt S^{13}\vt S^{23} & \vt S^{12}\vt S^{23}+\vt S^{13}\vt S^{22} & \vt S^{13}\vt S^{21}+\vt S^{11}\vt S^{23} & \vt S^{11}\vt S^{22}+\vt S^{12}\vt S^{21} 
\end{pmatrix}.
\end{equation}
\end{strip}

This means that if we want to compute the six components on $\mathcal Z$ of a strain that has components equal to $\ee$ on the \emph{transformed} basis $(\tilde{\Z}_k)_{k=1}^6$, we just need to compute $\hat{S}\ee$.
Similarly, the vector $\hat{S}\ttau$ gives the components on $\mathcal Z$ of a stress that has components equal to $\ttau$ on the \emph{transformed} basis $(\tilde{\Z}_k)_{k=1}^6$.
Note that there is no one-to-one correspondence between isometries and the set of induced transformations, since different isometries of $\R^3$ can be associated with the same $\hat{S}$, as we shall see below. This is due to the fact that different isometries acting on the lattice directors may induce the same transformation on the elements of the tensorial basis $\mathcal Z$.

A symmetry of the lattice $\mathcal L$ is a purely geometric concept.
We will now discuss how these isometries can be related to the material response and how they influence the large-scale mechanics of the homogenized system.
An isometry $\vt S$ of $\R^3$ is a \emph{material symmetry} if the stress computed on the transformed strain represented by $\hat{S}\ee$, namely $C\hat{S}\ee$, coincides with the transformed stress $\hat{S}C\ee$ for any choice of $\ee$. Hence, requiring the stress identity $C\hat{S}\ee=\hat{S}C\ee$ corresponds, by the arbitrariness of the strain, to the commutation relation 
\begin{equation}\label{eq:commutation}
C\hat{S}=\hat{S}C
\end{equation}
between the matrix representing the isometry and the elasticity matrix.

If the inclusion lattice possesses a given symmetry $\vt S$, the materials are isotropic, and the inclusions are spherical, then it is clear that the strains $\ee$ and $\hat{S}\ee$ represent equivalent stimuli on the material, because the two situations can be made \emph{completely} identical by a mere change of coordinates. The full experiment is coherently rotated and so are the reactions of each material region. In this case we expect that every symmetry of the lattice is linked to a corresponding material symmetry.

Nevertheless, we stress that for a system with a given inclusion lattice the group $\gm$ of material symmetries can be larger than the group $\gl$ of lattice symmetries.
In fact, under the present assumptions of isotropy and spherical inclusions, any lattice symmetry is a material symmetry and we conclude that equation~\eqref{eq:commutation} must be satisfied for $\hat{S}$ associated with elements of $\gl$. 
This imposes constraints in the form of linear relations between the components of the elasticity tensor $C$.
Once these constraints are identified, we may find additional transformations $\hat{S}$ that satisfy \eqref{eq:commutation} for any \emph{constrained} $C$ but are \emph{not} associated to any lattice symmetry.
If this is the case, then $\gm$ is strictly larger than $\gl$.
As shown in Section~\ref{sec:sympcell}, finding lattices with this property is the key point of the present treatment.

\subsection{Examples of material symmetries}

The isometries given by the identity $\I_3$ of $\R^3$ and by $-\I_3$ (that are symmetries of any periodic lattice) both induce the identity transformation $\hat{S}=I_6$ and the commutation relation $I_6C=CI_6$ is true for any $C$. Hence, if the lattice and material symmetry groups are $\gl=\gm=\{\I_3,-\I_3\}$, no constraint is imposed on $C$ and the material is generically anisotropic. 
In what follows, we denote rotations by $\vt{Q}$ and reflections by $\vt{R}$, with explanatory subscripts.

Let us now consider a lattice that is invariant under $\I$, $-\I$, and under rotations of angle $\pi$ about one of the directors, say $\vc l_3$. 
This is the case whenever the lattice generator $\vc a_3$ is orthogonal to the plane identified by $\vc a_1$ and $\vc a_2$, and thus coincides with $\vc l_3$.
The relevant isometry is given by
\[
\vt Q_\pi=\begin{pmatrix}
-1 & 0 & 0 \\
0 & -1 & 0 \\
 0 & 0 & 1
\end{pmatrix}.
\]
By composition, the lattice must also be invariant under $-\vt Q_\pi=-\I\vt Q_\pi$, that is a reflection through the plane generated by $\vc l_1$ and $\vc l_2$.
Since $\vt Q_\pi^2=(-\vt Q_\pi)^2=\I$, no other symmetry is implied by the group structure and $\gl=\{\I,-\I,\vt Q_\pi,-\vt Q_\pi\}$.
The transformation induced by both $\vt Q_\pi$ and $-\vt Q_\pi$ is
\[
\hat{Q}_\pi=\begin{pmatrix}
1 & 0 & 0 & 0 & 0 & 0 \\
0 & 1 & 0 & 0 & 0 & 0 \\
0 & 0 & 1 & 0 & 0 & 0 \\
0 & 0 & 0 & -1 & 0 & 0 \\
0 & 0 & 0 & 0 & -1 & 0 \\
0 & 0 & 0 & 0 & 0 & 1
\end{pmatrix}
\]
and the commutation relation $\hat{Q}_\pi C=C\hat{Q}_\pi$ implies $C_{4k}=C_{k4}=C_{5k}=C_{k5}=0$ for $k\neq 4,5$.
Since the diagonal entries of $C$ remain unconstrained, no additional transformation satisfies the commutation relation for all constrained elasticity matrices and we have $\gm=\gl=\{\I,-\I,\vt Q_\pi,-\vt Q_\pi\}$. 
In this case the large-scale response of the material is \emph{monoclinic}.

If it happens that there is invariance of the lattice under $\pi$-rotations about another axis $\vc b$ \emph{orthogonal to} $\vc l_3$, then we easily see that the last three columns and rows of $C$ have non-vanishing coefficients only on diagonal entries.
This is the case of an \emph{orthotropic} material.
Note that in this case the group of lattice symmetries contains also a third $\pi$-rotation about $\vc l_3\times\vc b$ and the corresponding reflection, for a total of eight elements, and we still have $\gm=\gl$ due to the independence of the diagonal entries of $C$.

Another important case arises when the material is orthotropic and also invariant under rotations of angle $\pi/2$ about, for instance, $\vc l_3$.
In this case we must consider the isometry
\[
\vt Q_\frac{\pi}{2}=\begin{pmatrix}
0 & -1 & 0 \\
 1 & 0 & 0 \\
 0 & 0 & 1
\end{pmatrix}.
\]
By composition we find that other isometries must belong to $\gl$. 
In particular, $\vt Q_\frac{3\pi}{2}=\vt Q^3_\frac{\pi}{2}$.
The transformation induced by $\vt Q_\frac{\pi}{2}$ is
\[
\hat{Q}_\frac{\pi}{2}=\begin{pmatrix}
0 & 1 & 0 & 0 & 0 & 0 \\
1 & 0 & 0 & 0 & 0 & 0 \\
0 & 0 & 1 & 0 & 0 & 0 \\
0 & 0 & 0 & 0 & 1 & 0 \\
0 & 0 & 0 & -1 & 0 & 0 \\
0 & 0 & 0 & 0 & 0 & -1
\end{pmatrix}.
\]
The commutation relation \eqref{eq:commutation} thus implies the additional constraints
\begin{equation}\label{eq:square}
\begin{aligned}
C_{21}=& C_{12},\quad  C_{22}=C_{11},\quad C_{23}=C_{13},\\ & C_{31}=C_{32},\quad C_{55}=C_{44}.
\end{aligned}
\end{equation}
This is the case for a lattice that is square in the plane generated by $\vc l_1$ and $\vc l_2$ and rectangular in the other two coordinate planes and the material response is \emph{tetragonal}.

If the lattice is simple cubic we have that $\pi/2$-rotations about $\vc l_1$ and $\vc l_2$ are additional symmetries.
The constraints imposed by the first of these can be obtained from \eqref{eq:square} with the index substitutions $3\mapsto 1$, $1\mapsto 2$, $2\mapsto 3$, $4\mapsto 5$, $5\mapsto 6$, and $6\mapsto 4$, leading to the cumulative identities
\begin{equation}\label{eq:cubic}
\begin{aligned}
&C_{33}=C_{22}=C_{11},\quad C_{66}=C_{55}=C_{44},\\ &C_{12}=C_{13}=C_{23}=C_{32}=C_{31}=C_{21}, 
\end{aligned}
\end{equation}
that already entail the consequences of the $\pi/2$-rotation about $\vc l_2$.
In this case the material response is \emph{cubic} and the elasticity matrix takes the form
\[
C_\mathrm{cubic}=\begin{pmatrix}
a & b & b & 0 & 0 & 0 \\
b & a & b & 0 & 0 & 0 \\
b & b & a & 0 & 0 & 0 \\
0 & 0 & 0 & c & 0 & 0 \\
0 & 0 & 0 & 0 & c & 0 \\
0 & 0 & 0 & 0 & 0 & c
\end{pmatrix},
\]
with $a$, $b$, and $c$ independent material constants.
Again, we have $\gm=\gl$.

\section{Symmetry-preserving periodic cells}\label{sec:sympcell}

We will now give a positive answer to the question \emph{``Are there inclusion lattices that would always preserve the material symmetries induced on the homogenized medium by the material symmetries of the two solid components and the shape of the inclusions?''}. 
In other words, we will exhibit an inclusion lattice for which the group of material symmetries induced by the lattice symmetries is the whole group of isometries of $\R^3$. In particular, for isotropic components and spherical inclusions, the large-scale elastic response of the material is isotropic.
Once an appropriate inclusion lattice is found, we can easily identify a rhomboidal cell that originates that lattice by a periodic tessellation of space.

We first treat the case of \emph{transverse isotropy}, that is isotropy in one plane.
We consider the periodic inclusion lattice with generators
\[
\vc a_1=(1,0,0),\quad \vc a_2=(1/2,\sqrt{3}/2,0),\quad \vc a_3=(0,0,1). 
\]
Clearly, we can choose as lattice directors $\vc l_1=\vc a_1$, $\vc l_3=\vc a_3$, and $\vc l_2=(0,1,0)$.  

The lattice formed in this way is hexagonal in planes orthogonal to $\vc l_3$.
The $\pi$-rotation $\vt Q_\pi$ about $\vc l_3$ is a material symmetry and so the system is at least monoclinic.
But in fact this lattice is also invariant under $\pi$-rotations about $\vc l_1$ and $\vc l_2$, showing that the material symmetry group contains that of an orthotropic system. 
We stress that these two symmetries are \emph{not} symmetries of the parallelepiped with edges identified by $(\vc a_1,\vc a_2,\vc a_3)$, which could be used as a periodic cell for computational studies of the homogenized response, and yet they belong to both the lattice and the material symmetry group.

A further symmetry characteristic of this lattice (but not of the periodic cell) is the $\pi/3$-rotation about $\vc l_3$:
\[
\vt Q_\frac{\pi}{3}=\begin{pmatrix}
1/2 & -\sqrt{3}/2 & 0 \\
 \sqrt{3}/2 & 1/2 & 0 \\
 0 & 0 & 1
\end{pmatrix}.
\]
The corresponding transformation is
\[
\hat{Q}_\frac{\pi}{3}=\begin{pmatrix}
1/4 & 3/4 & 0 & 0 & 0 & -\sqrt{6}/4 \\
3/4 & 1/4 & 0 & 0 & 0 & \sqrt{6}/4 \\
0 & 0 & 1 & 0 & 0 & 0 \\
0 & 0 & 0 & 1/2 & \sqrt{3}/2 & 0 \\
0 & 0 & 0 & -\sqrt{3}/2 & 1/2 & 0 \\
\sqrt{6}/4 & -\sqrt{6}/4 & 0 & 0 & 0 & -1/2  
\end{pmatrix}.
\]

Taking into account that the orthotropic symmetry implies that the last three rows and columns of the elasticity matrix $C$ have non-vanishing ciefficients only on diagonal entries, it can be readily seen that the commutation relation $\hat{Q}_\frac{\pi}{3}C=C\hat{Q}_\frac{\pi}{3}$ implies
\begin{equation}\label{eq:transverse-isotropy}
\begin{aligned}
& C_{21}=C_{12},
\quad C_{22}=C_{11},
\quad C_{13}=C_{23},\\
& C_{31}=C_{32},
\quad C_{55}=C_{44},\quad C_{11}=C_{12}+C_{66},
\end{aligned}
\end{equation}
entailing the following general form of the elasticity matrix $C_\mathrm{trans}$ associated with this inclusion lattice:
\[
C_\mathrm{trans}=\begin{pmatrix}
a & a-b & d & 0 & 0 & 0 \\
a-b & a & d & 0 & 0 & 0 \\
d' & d' & c & 0 & 0 & 0 \\
0 & 0 & 0 & e & 0 & 0 \\
0 & 0 & 0 & 0 & e & 0 \\
0 & 0 & 0 & 0 & 0 & b
\end{pmatrix},
\]
with $a$, $b$, $c$, $d$, $d'$, and $e$ are independent material constants, with $d=d'$ if we assume the existence of an elastic energy density.

An observation central to our argument is that we arrive at the form $C_\mathrm{trans}$ for the elasticity matrix by assuming, on top of an orthotropic symmetry group. the \emph{sole} addition of the $\pi/3$-rotation $\vt Q_\frac{\pi}{3}$ to the set of lattice and material symmetries.
Nevertheless, it is just a matter of simple computations to check that an elasticity tensor of the special form $C_\mathrm{trans}$ does in fact commute with any of the transformation matrices associated with rotations of arbitrary angle $\theta$ about $\vc l_3$:
\[
\vt Q_\theta=\begin{pmatrix}
\cos\theta & -\sin\theta & 0 \\
 \sin\theta & \cos\theta & 0 \\
 0 & 0 & 1
\end{pmatrix}.
\]
This shows that $\gm$ contains all the isometries that preserve the plane orthogonal to $\vc l_3$, it is strictly larger than $\gl$, and the material response is \emph{transversely isotropic}.

An analogous result was proven  in a more general setting regarding homogenized elastic structures by Ptashnyk and Seguin~\cite{Ptashnyk_2016}.
What we want to highlight here is the shape of the  computational cell, as defined by $(\vc a_1,\vc a_2,\vc a_3)$, that can be used in simulations and design processes to guarantee that transverse isotropy is not disrupted by the assumed periodicity (that is of course an approximation in the description of real systems).
The fact that hexagonal periodic cells preserve isotropy in planar problems has been empirically know for quite some time.
Nevertheless, from the discussion above, it should be clear that we do not need to have a periodic cell that is invariant under $\pi/3$-rotations as long as we generate a periodic inclusion lattice with this property.
In fact, we can use a cell with faces orthogonal to $\vc l_3$ that are rhombi, with angles of $60^\circ$ and $120^\circ$, normally extruded in the third direction. 

We are now positioned to make an important step beyond what has been so far rigorously or empirically shown.
Indeed, it is difficult to bring to three-dimensional cells a hexagon-like character, but we can easily tune a rhomboidal cell to achieve an inclusion lattice that implies the largest possible material symmetry group, namely the full group of isometries.
With this type of cell we can simulate or design periodic structures that do not disrupt any material symmetry in spite of the approximation associated with periodicity.
We provide a way other than randomization to generate isotropically elastic metamaterials. 

\begin{figure}
    \centering
    \includegraphics[width=7.2cm]{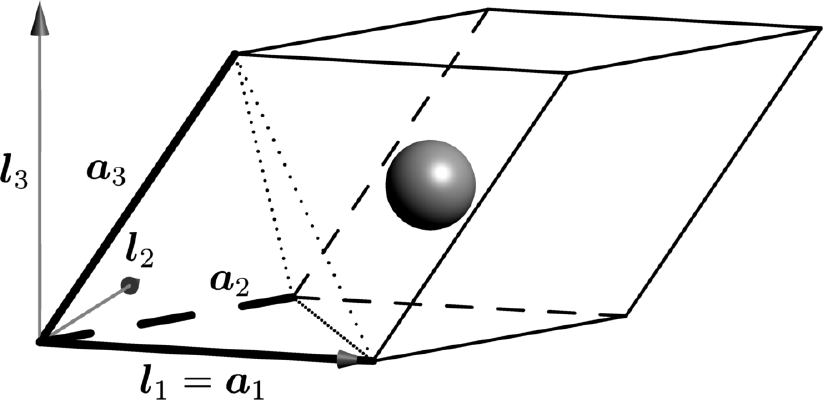}
    \caption{Unit cell that generates the FCC inclusion lattice. The generators $(\vc a_1,\vc a_2,\vc a_3)$ are edges of a regular tetrahedron (dotted lines) and the faces are identical rhombi with angles of $60^\circ$ and $120^\circ$.}
    \label{fig:fig1}
\end{figure}

We consider the periodic inclusion lattice of unit cell shown in Figure~\ref{fig:fig1} and with generators
\[
\vc a_1=(1,0,0),\; \vc a_2=\left(\frac12,\frac{\sqrt{3}}{2},0\right),\; \vc a_3=\left(\frac12, \frac{\sqrt{3}}{6}, \frac{\sqrt{6}}{3}\right),
\]
that are edges of a regular tetrahedron.
We then choose as lattice directors $\vc l_1=\vc a_1$, $\vc l_2=(0,1,0)$, and $\vc l_3=(0,0,1)$.  
In each of the planes generated by pairs of lattice generators, we have a hexagonal lattice, but the symmetries of these planar lattices need not be symmetries of the three-dimensional structure.
For instance the rotations of angle $\pi/3$ about $\vc l_3$ is not a lattice symmetry.

The first lattice symmetries that we consider are the reflections that map $\vc a_1$ onto $-\vc a_1$ and $\vc a_2$ onto $-\vc a_2$, namely
\[
\vt R_1=\begin{pmatrix}
 -1 & 0 & 0 \\
 0 & 1 & 0 \\
 0 & 0 & 1
 \end{pmatrix},\text{ and }
\vt R_2=\begin{pmatrix}
1/2 & -\sqrt{3}/2 & 0 \\
 -\sqrt{3}/2 & -1/2 & 0 \\
 0 & 0 & 1
\end{pmatrix}.
\] 
From the commutation relation originated by $\vt R_1$ we find the orthotropic symmetry with constraints $C_{5k}=C_{k5}=C_{6k}=C_{k6}=0$ for $k\neq 5,6$.
The additional consequences of the lattice symmetry $\vt R_2$ are
\begin{gather*}
C_{43}=C_{34}=0,\quad C_{22}=C_{11}, \quad C_{12}=C_{21}=C_{11}-C_{66},\\ C_{23}=C_{13}, \quad C_{32}=C_{31},\quad
C_{24}=-C_{14}, \quad C_{42}=-C_{41}, \\ C_{55}=C_{44}, \quad C_{56}=\sqrt{2}C_{41}, \quad C_{65}=\sqrt{2}C_{14}.
\end{gather*}
With these constraints, the elasticity matrix can be given in terms of 8 parameters, with $D=C_{11}-C_{66}$, as
\begin{equation}\label{eq:8par}
\left(
\begin{array}{cccccc}
 C_{11} & D & C_{13} & C_{14} & 0 & 0 \\
 D & C_{11} & C_{13} & -C_{14} & 0 & 0 \\
 C_{31} & C_{31} & C_{33} & 0 & 0 & 0 \\
 C_{41} & -C_{41} & 0 & C_{44} & 0 & 0 \\
 0 & 0 & 0 & 0 & C_{44} & \sqrt{2} C_{41} \\
 0 & 0 & 0 & 0 & \sqrt{2} C_{14} & C_{66} \\
\end{array}
\right),
\end{equation}
that shows some signatures of isotropic symmetry but is not even transversely isotropic yet.

The last set of constraints that we need to consider comes from the $\pi/3$ rotation about the axis identified by $\vc a_1+\vc a_2+\vc a_3$, a diagonal of the rhomboidal cell.
This produces a cyclic permutation of the generators and corresponds to
\[
\vt Q_\mathrm{sum}=\frac{1}{4}
\left(
\begin{array}{ccc}
 1 & \sqrt{3}+2 & \sqrt{2}-\sqrt{6} \\
 \sqrt{3}-2 & -1 & -\sqrt{2}-\sqrt{6} \\
 -\sqrt{2}-\sqrt{6} & \sqrt{6}-\sqrt{2} & 0 \\
\end{array}
\right).
\]
The linear transformation associated with this rotation is

\begin{strip}
\[
\hat{Q}_\mathrm{sum}=\frac{1}{16}
\left(
\begin{array}{cccccc}
 1 & 4 \sqrt{3}+7 & 8-4 \sqrt{3} & -2 \sqrt{3}-2 & 2-2 \sqrt{3} & 2 \sqrt{2}+\sqrt{6} \\
 7-4 \sqrt{3} & 1 & 4 \sqrt{3}+8 & 2 \sqrt{3}+2 & 2 \sqrt{3}-2 & 2 \sqrt{2}-\sqrt{6} \\
 4 \sqrt{3}+8 & 8-4 \sqrt{3} & 0 & 0 & 0 & -4 \sqrt{2} \\
 2 \sqrt{3}-2 & 2-2 \sqrt{3} & 0 & -4 & 4 \sqrt{3}+8 & 6 \sqrt{2}-2 \sqrt{6} \\
 -2 \sqrt{3}-2 & 2 \sqrt{3}+2 & 0 & 4 \sqrt{3}-8 & 4 & -6 \sqrt{2}-2 \sqrt{6} \\
 \sqrt{6}-2 \sqrt{2} & -2 \sqrt{2}-\sqrt{6} & 4 \sqrt{2} & -6 \sqrt{2}-2 \sqrt{6} & 2 \sqrt{6}-6 \sqrt{2} & -2 \\
\end{array}
\right)
\]
\end{strip}

From the commutation relation $C\hat{Q}_\mathrm{sum}=\hat{Q}_\mathrm{sum}C$, with $C$ as given in equation~\eqref{eq:8par}, and by neglecting redundant identities, we finally obtain for $C$ the constraints
\begin{gather*}
C_{33}=C_{11}, \quad C_{13}=C_{31}=C_{11}-C_{66}, \\ C_{14}=C_{41}=0, \quad C_{44}=C_{66}.
\end{gather*} 
The elasticity matrix thus depends only on two independent material constants, denoted by $a$ and $b$, and assumes the general form
\[
C_\mathrm{iso}=\begin{pmatrix}
a & a-b & a-b & 0 & 0 & 0 \\
a-b & a & a-b & 0 & 0 & 0 \\
a-b & a-b & a & 0 & 0 & 0 \\
0 & 0 & 0 & b & 0 & 0 \\
0 & 0 & 0 & 0 & b & 0 \\
0 & 0 & 0 & 0 & 0 & b
\end{pmatrix}.
\]

This is the form of the elasticity matrix associated with an isotropic material response.
In fact, even though we arrived at $C_\mathrm{iso}$ by imposing the commutation relation \eqref{eq:commutation} for only three different material symmetries, it can be easily checked that $C_\mathrm{iso}\hat{S}=\hat{S}C_\mathrm{iso}$ for \emph{any} $\hat{S}$ of the form given by \eqref{eq:hatS}, namely any isometry of $\R^3$ is a material symmetry.

\section{Conclusions}

We have proved that a periodic arrangement on a Face-Centered Cubic lattice of spherical inclusions of an isotropic solid within an isotropic matrix gives rise to a large-scale isotropic response.
This has the important implication that a rhomboidal computational cell that generates such a lattice can be used to design composites in which the material symmetry is not affected by the periodicity of the construction, since even the largest possible symmetry group would be preserved.

To be able to discuss material symmetries in a way that is independent of the reference frame, we have introduced a normalized Voigt representation, based on material directors rather than on a coordinate basis.
Within this concise setup, symmetries of the inclusion lattice induce linear constraints on the entries of the 6 by 6 matrix of material coefficients that represents the linear elasticity tensor.
Importantly, such geometric constraints can give rise to a material symmetry group which is larger than the group of lattice symmetries, thereby leaving room for the emergence of effectively isotropic materials.

Our findings allow to interpret several computational results that show an almost isotropic behavior of metamaterials with FCC structure under a new light. 
Those results should be regarded as missing the theoretical prediction simply by numerical approximation and not as showing that the metamaterial has cubic symmetry with a small, but non-vanishing, degree of anisotropy.

Our construction is based on the remarkable fact that a rather small number of discrete symmetries is sufficient to constrain the elasticity tensor in such a way that any isometry becomes a material symmetry.
In light of this, the shape of the inclusions need not be spherical. It is sufficient to choose a solid that is symmetric under the two reflections $\vt R_1$ and $\vt R_2$ and the rotation $\vt Q_\mathrm{sum}$ considered above.
For instance, an appropriately oriented tetrahedron would be suitable, but also more complicated shapes could be considered.

\section*{Acknowledgments}
RP is partially supported by EPSRC research grants EP/S030875/1 and EP/T017899/1 and conducted the research according to the inspiring scientific principles of the national Italian mathematics association INdAM (``Istituto nazionale di Alta Matematica”).
GGG acknowledges the support of the National Group of Mathematical Physics (GNFM--INdAM) through the funding scheme ``GNFM Young Researchers’ Projects 2020''.


\begin{thebibliography}{26}


\providecommand{\url}[1]{\texttt{#1}}
\expandafter\ifx\csname urlstyle\endcsname\relax
  \providecommand{\doi}[1]{doi: #1}\else
  \providecommand{\doi}{doi: \begingroup \urlstyle{rm}\Url}\fi

\bibitem[1]{cherkaev1997topics}
\textsc{Cherkaev}, Andrej ; \textsc{Kohn}, Robert:
\newblock \emph{Topics in the mathematical modelling of composite materials}.
\newblock Springer, 1997.
\newblock \url{http://dx.doi.org/10.1007/978-1-4612-2032-9}.
\newblock \url{http://dx.doi.org/10.1007/978-1-4612-2032-9}

\bibitem[2]{milton2002theory}
\textsc{Milton}, Graeme~W.:
\newblock \emph{The theory of composites}. Bd.~6.
\newblock Cambridge University Press, 2002.
\newblock \url{http://dx.doi.org/10.1017/CBO9780511613357}.
\newblock \url{http://dx.doi.org/10.1017/CBO9780511613357}

\bibitem[3]{hull1996introduction}
\textsc{Hull}, Derek ; \textsc{Clyne}, TW:
\newblock \emph{An introduction to composite materials}.
\newblock Cambridge university press, 1996.
\newblock \url{http://dx.doi.org/10.1017/CBO9781139170130}.
\newblock \url{http://dx.doi.org/10.1017/CBO9781139170130}

\bibitem[4]{nicholas2020integrating}
\textsc{Nicholas}, Paul ; \textsc{Rossi}, Gabriella ; \textsc{Williams}, Ella ;
  \textsc{Bennett}, Michael  ; \textsc{Schork}, Tim:
\newblock Integrating real-time multi-resolution scanning and machine learning
  for Conformal Robotic 3D Printing in Architecture.
\newblock {In: }\emph{International Journal of Architectural Computing} 18
  (2020), Nr. 4, S. 371--384.
\newblock \url{http://dx.doi.org/10.1177/1478077120948203}. --
\newblock DOI 10.1177/1478077120948203

\bibitem[5]{suresh2020review}
\textsc{Suresh~Kumar}, N ; \textsc{Padma~Suvarna}, R ; \textsc{Chandra
  Babu~Naidu}, K ; \textsc{Banerjee}, Prasun ; \textsc{Ratnamala}, A  ;
  \textsc{Manjunatha}, H:
\newblock A review on biological and biomimetic materials and their
  applications.
\newblock {In: }\emph{Applied Physics A} 126 (2020), Nr. 6, S. 1--18.
\newblock \url{http://dx.doi.org/10.1007/s00339-020-03633-z}. --
\newblock DOI 10.1007/s00339--020--03633--z

\bibitem[6]{HoriNematNasser1999}
\textsc{Hori}, Muneo ; \textsc{Nemat-Nasser}, Sia:
\newblock On two micromechanics theories for determining micro-macro relations
  in heterogeneous solids.
\newblock {In: }\emph{Mechanics of Materials} 31 (1999), Nr. 10, S. 667--682.
\newblock \url{http://dx.doi.org/10.1016/S0167-6636(99)00020-4}. --
\newblock DOI 10.1016/S0167--6636(99)00020--4

\bibitem[7]{royer2019quasi}
\textsc{Royer}, Pascale ; \textsc{Recho}, Pierre  ; \textsc{Verdier}, Claude:
\newblock On the quasi-static effective behaviour of poroelastic media
  containing elastic inclusions.
\newblock {In: }\emph{Mechanics Research Communications} 96 (2019), S. 19--23.
\newblock \url{http://dx.doi.org/10.1016/j.mechrescom.2019.02.004}. --
\newblock DOI 10.1016/j.mechrescom.2019.02.004

\bibitem[8]{penta2016can}
\textsc{Penta}, R ; \textsc{Raum}, K ; \textsc{Grimal}, Q ; \textsc{Schrof}, S
  ; \textsc{Gerisch}, A:
\newblock Can a continuous mineral foam explain the stiffening of aged bone
  tissue? A micromechanical approach to mineral fusion in musculoskeletal
  tissues.
\newblock {In: }\emph{Bioinspiration \& biomimetics} 11 (2016), Nr. 3, S.
  035004.
\newblock \url{http://dx.doi.org/10.1088/1748-3190/11/3/035004}. --
\newblock DOI 10.1088/1748--3190/11/3/035004

\bibitem[9]{collis2017multi}
\textsc{Collis}, Joe ; \textsc{Hubbard}, Matthew~E.  ; \textsc{O'Dea},
  Reuben~D.:
\newblock A multi-scale analysis of drug transport and response for a
  multi-phase tumour model.
\newblock {In: }\emph{European Journal of Applied Mathematics} 28 (2017), Nr.
  3, S. 499--534.
\newblock \url{http://dx.doi.org/10.1017/S0956792516000413}. --
\newblock DOI 10.1017/S0956792516000413

\bibitem[10]{nika2019design}
\textsc{Nika}, Grigor ; \textsc{Constantinescu}, Andrei:
\newblock Design of multi-layer materials using inverse homogenization and a
  level set method.
\newblock {In: }\emph{Computer Methods in Applied Mechanics and Engineering}
  346 (2019), S. 388--409.
\newblock \url{http://dx.doi.org/10.1016/j.cma.2018.11.029}. --
\newblock DOI 10.1016/j.cma.2018.11.029

\bibitem[11]{Berger_2017}
\textsc{Berger}, JB ; \textsc{Wadley}, HNG  ; \textsc{McMeeking}, RM:
\newblock Mechanical metamaterials at the theoretical limit of isotropic
  elastic stiffness.
\newblock {In: }\emph{Nature} 543 (2017), Nr. 7646, S. 533--537.
\newblock \url{http://dx.doi.org/10.1038/nature21075}. --
\newblock DOI 10.1038/nature21075

\bibitem[12]{Rossi_2020}
\textsc{Rossi}, N ; \textsc{Yera}, Rolando ; \textsc{M{\'e}ndez}, CG ;
  \textsc{Toro}, Sebastian  ; \textsc{Huespe}, Alfredo~E.:
\newblock Numerical technique for the 3D microarchitecture design of elastic
  composites inspired by crystal symmetries.
\newblock {In: }\emph{Computer Methods in Applied Mechanics and Engineering}
  359 (2020), S. 112760.
\newblock \url{http://dx.doi.org/10.1016/j.cma.2019.112760}. --
\newblock DOI 10.1016/j.cma.2019.112760

\bibitem[13]{Yera_2020}
\textsc{Yera}, Rolando ; \textsc{Rossi}, N ; \textsc{Mendez}, CG  ;
  \textsc{Huespe}, Alfredo~E.:
\newblock Topology design of 2D and 3D elastic material microarchitectures with
  crystal symmetries displaying isotropic properties close to their theoretical
  limits.
\newblock {In: }\emph{Applied Materials Today} 18 (2020), S. 100456.
\newblock \url{http://dx.doi.org/10.1016/j.apmt.2019.100456}. --
\newblock DOI 10.1016/j.apmt.2019.100456

\bibitem[14]{Wu_2021}
\textsc{Wu}, Jun ; \textsc{Sigmund}, Ole  ; \textsc{Groen}, Jeroen~P.:
\newblock Topology optimization of multi-scale structures: a review.
\newblock {In: }\emph{Structural and Multidisciplinary Optimization}  (2021),
  S. 1--26.
\newblock \url{http://dx.doi.org/10.1007/s00158-021-02881-8}. --
\newblock DOI 10.1007/s00158--021--02881--8

\bibitem[15]{Rossi_2021}
\textsc{Rossi}, Nestor ; \textsc{Podest{\'a}}, Juan~M. ; \textsc{Bre}, Facundo
  ; \textsc{M{\'e}ndez}, Carlos~G.  ; \textsc{Huespe}, Alfredo~E.:
\newblock A microarchitecture design methodology to achieve extreme isotropic
  elastic properties of composites based on crystal symmetries.
\newblock {In: }\emph{Structural and Multidisciplinary Optimization} 63 (2021),
  Nr. 5, S. 2459--2472.
\newblock \url{http://dx.doi.org/10.1007/s00158-020-02823-w}. --
\newblock DOI 10.1007/s00158--020--02823--w

\bibitem[16]{Forte_1996}
\textsc{Forte}, Sandra ; \textsc{Vianello}, Maurizio:
\newblock Symmetry classes for elasticity tensors.
\newblock {In: }\emph{Journal of Elasticity} 43 (1996), Nr. 2, S. 81--108.
\newblock \url{http://dx.doi.org/10.1007/BF00042505}. --
\newblock DOI 10.1007/BF00042505

\bibitem[17]{Bona_2004}
\textsc{Bona}, Andrej ; \textsc{Bucataru}, Ioan  ; \textsc{Slawinski},
  Michael~A.:
\newblock Material symmetries of elasticity tensors.
\newblock {In: }\emph{Quarterly Journal of Mechanics and Applied Mathematics}
  57 (2004), Nr. 4, S. 583--598.
\newblock \url{http://dx.doi.org/10.1093/qjmam/57.4.583}. --
\newblock DOI 10.1093/qjmam/57.4.583

\bibitem[18]{Ptashnyk_2016}
\textsc{Ptashnyk}, Mariya ; \textsc{Seguin}, Brian:
\newblock Periodic homogenization and material symmetry in linear elasticity.
\newblock {In: }\emph{Journal of Elasticity} 124 (2016), Nr. 2, S. 225--241.
\newblock \url{http://dx.doi.org/10.1007/s10659-015-9566-x}. --
\newblock DOI 10.1007/s10659--015--9566--x

\bibitem[19]{Podesta_2019}
\textsc{Podest{\'a}}, Juan~M. ; \textsc{M{\'e}ndez}, CM ; \textsc{Toro},
  Sebastian  ; \textsc{Huespe}, Alfredo~E.:
\newblock Symmetry considerations for topology design in the elastic inverse
  homogenization problem.
\newblock {In: }\emph{Journal of the Mechanics and Physics of Solids} 128
  (2019), S. 54--78.
\newblock \url{http://dx.doi.org/10.1016/j.jmps.2019.03.018}. --
\newblock DOI 10.1016/j.jmps.2019.03.018

\bibitem[20]{Mendez_2019}
\textsc{M{\'e}ndez}, C ; \textsc{Podest{\'a}}, JM ; \textsc{Toro}, S ;
  \textsc{Huespe}, Alfredo~E.  ; \textsc{Oliver}, Javier:
\newblock Making use of symmetries in the three-dimensional elastic inverse
  homogenization problem.
\newblock {In: }\emph{International Journal for Multiscale Computational
  Engineering} 17 (2019), Nr. 3.
\newblock \url{http://dx.doi.org/10.1615/IntJMultCompEng.2019029111}. --
\newblock DOI 10.1615/IntJMultCompEng.2019029111

\bibitem[21]{sabina2002overall}
\textsc{Sabina}, Federico~J. ; \textsc{Bravo-Castillero}, Juli{\'a}n ;
  \textsc{Guinovart-D{\'i}az}, Ra{\'u}l ; \textsc{Rodr{\'i}guez-Ramos},
  Reinaldo  ; \textsc{Valdiviezo-Mijangos}, Oscar~C.:
\newblock Overall behavior of two-dimensional periodic composites.
\newblock {In: }\emph{International journal of solids and structures} 39
  (2002), Nr. 2, S. 483--497.
\newblock \url{http://dx.doi.org/10.1016/S0020-7683(01)00107-X}. --
\newblock DOI 10.1016/S0020--7683(01)00107--X

\bibitem[22]{parnell2006dynamic}
\textsc{Parnell}, William~J. ; \textsc{Abrahams}, I~D.:
\newblock Dynamic homogenization in periodic fibre reinforced media.
  Quasi-static limit for SH waves.
\newblock {In: }\emph{Wave Motion} 43 (2006), Nr. 6, S. 474--498.
\newblock \url{http://dx.doi.org/10.1016/j.wavemoti.2006.03.003}. --
\newblock DOI 10.1016/j.wavemoti.2006.03.003

\bibitem[23]{penta2015investigation}
\textsc{Penta}, Raimondo ; \textsc{Gerisch}, Alf:
\newblock Investigation of the potential of asymptotic homogenization for
  elastic composites via a three-dimensional computational study.
\newblock {In: }\emph{Computing and Visualization in Science} 17 (2015), Nr. 4,
  S. 185--201.
\newblock \url{http://dx.doi.org/10.1007/s00791-015-0257-8}. --
\newblock DOI 10.1007/s00791--015--0257--8

\bibitem[24]{penta2017asymptotic}
\textsc{Penta}, Raimondo ; \textsc{Gerisch}, Alf:
\newblock The asymptotic homogenization elasticity tensor properties for
  composites with material discontinuities.
\newblock {In: }\emph{Continuum Mechanics and Thermodynamics} 29 (2017), Nr. 1,
  S. 187--206.
\newblock \url{http://dx.doi.org/10.1007/s00161-016-0526-x}. --
\newblock DOI 10.1007/s00161--016--0526--x

\bibitem[25]{Steigmann_2007}
\textsc{Steigmann}, David~J.:
\newblock On the frame invariance of linear elasticity theory.
\newblock {In: }\emph{Zeitschrift f{\"u}r angewandte Mathematik und Physik} 58
  (2007), Nr. 1, S. 121--136.
\newblock \url{http://dx.doi.org/10.1007/s00033-006-6047-x}. --
\newblock DOI 10.1007/s00033--006--6047--x

\end{thebibliography}
\end{document}